# Studies of Optical Damage in Photorefractive Single LiNbO$_3$ Crystals using Imaging Polarimetry

Krupych O., Smaga I. and Vlokh R.

Institute of Physical Optics, 23 Dragomanov St., 79005 Lviv, Ukraine: vlokh@ifo.lviv.ua

## Abstract

The optical damage of photorefractive material, single LiNbO$_3$ crystal, is experimentally studied. The specimen has been illuminated with the radiation of continuous Ar-laser (the wavelength of 488 nm) focused to 35μm spot. The induced birefringence map is obtained by means of imaging polarimeter. Promising resources of the experimental setup for detecting laser-induced damage in photorefractive materials is demonstrated.

## Introduction

It is known that photorefractive crystals can change refractive indices when illuminated with continuous lasers [1,2]. The phenomenon is often called as an optical damage of photorefractive crystals. It is clear that such the damage is quite different from a common laser damage related to optical breakdown of solids under the action of high-power laser radiation. The latter one leads to irreversible destructive changes in the media under illumination [3,4,5,6]. In case of refractive crystals, optically induced changes in the refractive indices are not connected with any destruction of material, although they restrict significantly the field of application of those materials in nonlinear and quantum optical devices. Notice also that heating of samples or their uniform illumination with UV radiation allow to return the samples to their initial homogeneous state.

To study optical damage in photorefractive media, optical polarization microscopy is usually used [2,7]. Unfortunately, this technique is laborious and time-consuming though it yields "rough" results. Therefore, the aim of this work is to prove a potential of imaging polarimetry in the studies of optical damage in photorefractive materials.

## Experimental results and discussion

While studying the optical damage of photorefractive crystals, we used the *X*-cut lithium niobate crystal (LiNbO$_3$ the thickness of 0.69 mm). We exploited the imaging polarimetric setup described earlier (see [8] and Fig. 1), with only minor recent technical modifications. Since the specimen was illuminated in the direction normal to the optic axis, it possessed optical birefringence, while the optical indicatrix orientation was invariable. Therefore, we removed the quarter-wave plate (the component 9 in Fig. 1) from the polarization generator and used linearly polarized incident light, with the azimuth 45° with respect to the optic axis direction. In this case, the output light generally is elliptically polarized, with the azimuth of 45° and the ellipticity expressed as

$$|\sin 2\varepsilon_o| = |\sin \Gamma| \quad \Rightarrow \quad |\varepsilon_o| = \left|\frac{\Gamma}{2}\right|, \quad (1)$$

where $\varepsilon_o$ is the ellipticity angle for the output light and $\Gamma$ the phase retardation in the specimen. Hence, after scanning the analyzer azimuth $a$, detecting the intensity $I$ and discriminating its background, one obtains the modulated signal described as follows:

$$I = \frac{I_0}{2} \cdot \left(1 + \cos 2\varepsilon_o \cos 2(a - \theta_o)\right), \quad (2)$$

where $I_0$ means the intensity of light passed through the specimen, $a$ the analyzer azimuth and $\theta_o$ the major axis azimuth of the polarization ellipse at the output of specimen.

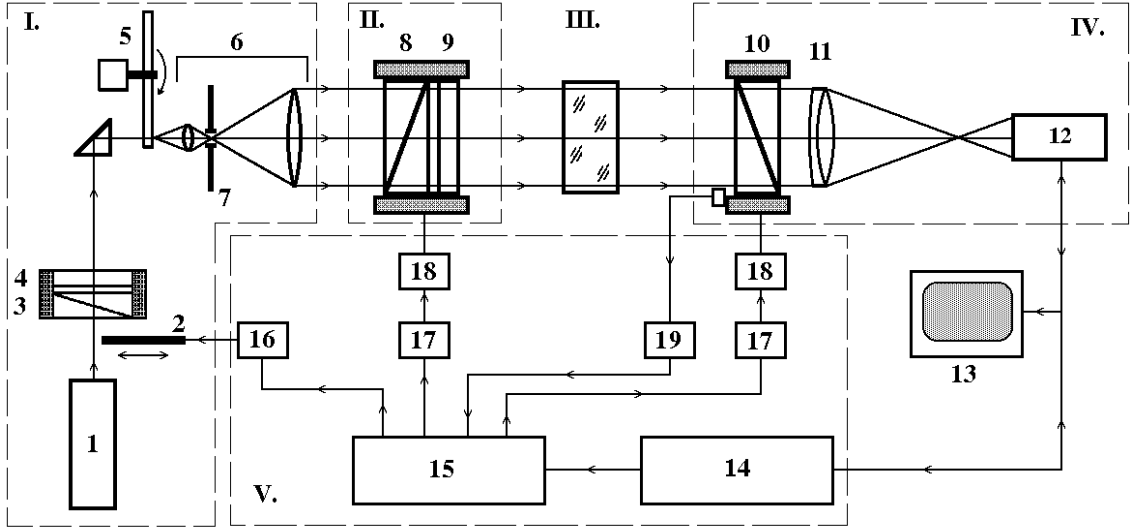

**Fig. 1.** Schematic representation of imaging polarimeter (I – light source section; II – polarization generator; III – specimen section; IV – polarization analyzer; and V – controlling unit): 1 – He-Ne laser; 2 – ray shutter; 3, 8 - polarizers; 4, 9 – quarter-wave plates; 5 – coherence scrambler; 6 – beam expander; 7 – spatial filter; 10 – analyzer; 11 – objective lens; 12 – CCD camera; 13 – TV monitor; 14 – framegrabber; 15 – PC; 16 – shutter's controller; 17 – step motors' controllers; 18 – step motors; and 19 – reference position controller.

Fitting this signal, we evaluate the $\varepsilon_o$ parameter and, consequently, the retardation $\Gamma$. Repeating the just described procedure of calculating $\Gamma$ for each pixel of the specimen image, we derive a map of optical retardation in the specimen.

First of all, we have fixed the retardation map characteristic of the specimen in its initial state. Then we have illuminated the specimen using a continuous Ar-laser (the intensity 95 mW and the wavelength 488 nm). The laser beam has been focused to the spot with the waist diameter of about 35 μm, corresponding to $1/e^2$ intensity level. The specimen has been kept in the waist position for 5 minutes. Then we have recorded the retardation map once again. The difference between these two maps gives us spatial distribution of the induced retardation $\delta\Gamma$ presented in Fig. 2. The induced birefringence itself may be calculated using the above pattern and the formula

$$\delta(\Delta n) = \frac{\delta\Gamma}{360°} \cdot \frac{\lambda}{t}, \qquad (3)$$

where $\delta(\Delta n)$ denotes the induced birefringence, $\lambda$ the light wavelength and $t$ the specimen thickness.

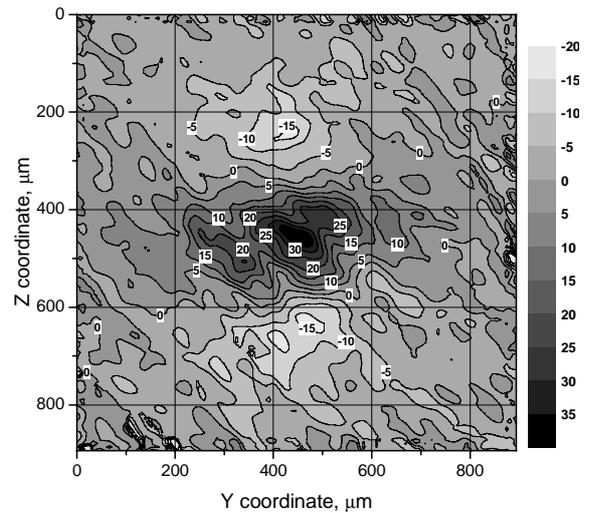

**Fig. 2.** Map of the induced retardation $\delta\Gamma$ (in arc degrees).

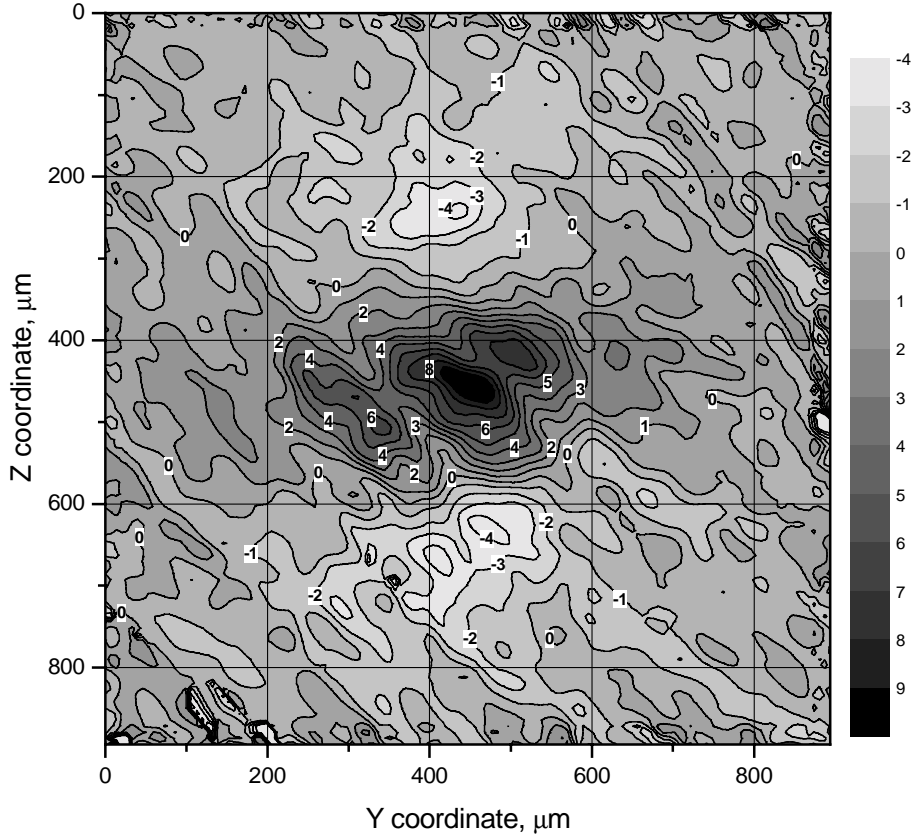

**Fig. 3.** Map of the induced birefringence $\delta(\Delta n)$ (the scaling coefficient is equal to $10^{-5}$; see also Fig. 4 and 5).

One can see from Fig. 3 that the spatial dependences of the induced birefringence differ notably for the directions of optic axis (the $Z$ coordinate) and its normal (the $Y$ coordinate). The sections of the surface of induced birefringence by the planes parallel and perpendicular to the optic axis, which refer to the centre of the illumination spot, are presented in Fig. 4 and 5, respectively. The $Z$ dependence of the induced birefringence exhibits positive changes in the birefringence occurring in the centre of the illumination spot and negative ones in peripheral regions. The $Y$ dependence has a bell-shaped character, manifesting no sign change in the induced birefringence.

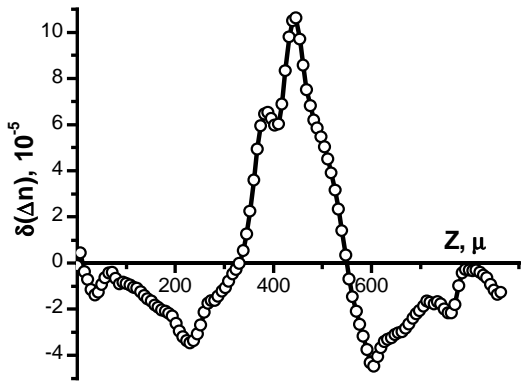

**Fig. 4.** Dependence of the induced birefringence on $Z$ coordinate.

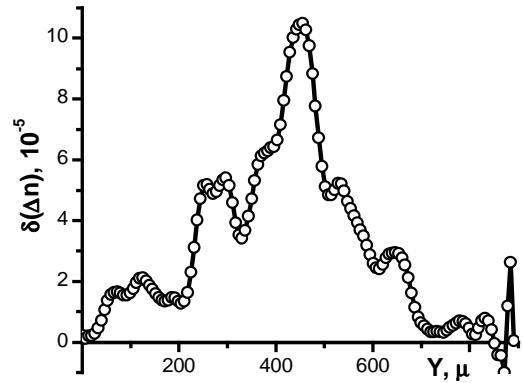

**Fig. 5.** Dependence of the induced birefringence on $Y$ coordinate.

As stated by *F.S. Chen* [2], the effect of optically induced change in the refractive indices is attributed to drift of photoexcited electrons out of the illuminated region, followed by their re-trapping near the beam periphery. The space-charge field between these re-trapped electrons and positive ionized centres in the illuminated region causes the observed changes in the refractive indices via electro-optic effect. In our case, the photoexcited electrons drift in the field of spontaneous polarization directed along the Z axis. The space charge formed this way has a shape of volume dipole. The direction of the electric field induced by the space charge is shown schematically in Fig. 6. It is clear that the directions of the induced electric fields are opposite in the regions ① and ② (see Fig. 6). This results in the opposite changes of the induced birefringence in these regions.

The maps of the induced retardation and the birefringence obtained by us indicate that the imaging polarimetric setup has a sufficient sensitivity and enables detecting and studying in detail the optical damage in photorefractive materials. The minimal changes detected in the induced retardation are estimated as ~ 5 deg and the same for the induced birefringence is about $10^{-5}$. Using common CCD cameras and objective lens allows achieving the image resolution of the order of magnitude of 10-50 μm. A possibility for varying optical magnification gives additional instrumental flexibility. As the procedures of measurements, image collecting and data refining are performed automatically under the software control, our technique has numerous advantages when compare with traditional optical microscopy.

## Conclusion

Our experimental results testify that the imaging polarimetry could be successfully used for detecting and studying quantitatively the optical damage in photorefractive media. The imaging polarimeter constructed at the Institute of Physical Optics has sufficient flexibility, sensitivity and spatial resolution. Al of these resources have been demonstrated on the instructive example of constructing the map of optically induced birefringence in the model single $LiNbO_3$ crystal.

## Acknowledgement

The authors are grateful to the Ministry of Education and Science of Ukraine (the Project N0104U000460) for financial support of this study.

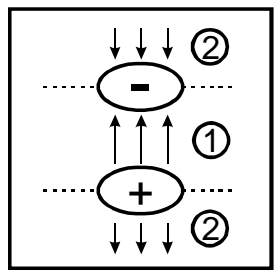

**Fig. 6.** Scheme of space charge and direction of the induced electric field.